\documentstyle[psfig,conf_iap]{article}
\begin{document}

\heading{%
%
Metal Enrichment of Ly alpha Clouds and Intergalactic Medium
%
} 
\par\medskip\noindent
\author{%
Izumi Murakami$^1$, Kazuyuki Yamashita$^2$
}
\address{%
National Institute for Fusion Science, Oroshi-cho, Toki, Gifu 509-52, 
Japan 
}
\address{%
Information Processing Center,
Chiba University,  Inage-ku, Chiba 263, Japan
}
%

\begin{abstract}
We have examined the metal enrichment of the intergalactic medium (IGM)
based on a galactic wind model.
A galactic wind driven by supernovae brings metallic gas to
the IGM but not so far beyond the gravitational potential.
The expanding velocity of the outflow depends on the star formation
timescale.
Examining 3D calculation for the IGM in CDM model, we find that
only 10 \% region  has metallicity larger than $10^{-2}Z_{\odot}$
at $z=3$.
Wide range of the IGM metallicity produces variety of CIV column densities
for a fixed HI column density.

\end{abstract}

\section{Introduction}

Recent observations have shown us that many Ly $\alpha$ clouds
are contaminated with metals.
Associated CIV absorption lines indicate that these clouds are likely to have
about 1/100 solar metallicity
\cite{Ty95}, \cite{CS95}, \cite{SC96}.
This puzzles us how such intergalactic clouds are metal-enriched,
since metals must be synthesized in stars, i.e. in galaxies.
Another question is whether Ly $\alpha$ clouds with lower
HI column densities are metal-enriched or not.

In this work we examine galactic wind 
driven by supernova explosions
to provide metals to the IGM,
in order to find answers for above questions.

\section{MODELS}

We have examined three models to investigate how a galactic wind
propagate and bring metallic gas to the IGM on the minihalo model:
a spherical cloud model,
a grid toy model,
and a 3D CDM model.

Assumptions over these models are summarized as follows:
Gas and dark matter are in a flat universe ($\Omega=1$)
with $H_0 = 50 \rm km \ s^{-1} Mpc^{-1}$ ($h=0.5$).
Stars are formed at high density and low temperature region.
Eleven percent of stars in  mass  become supernovae 
with releasing
$10^{51}$erg and $3.2 M_{\odot}$ metallic gas per an average supernova
which mass is $20.5 M_{\odot}$.
Uniform UV background radiation is assumed with 
power-law spectrum of $\alpha=1$. 
The flux evolution model is assumed 
similarly to Haardt \& Madau's model \cite{HM}
and
$J_{21}= 0.72$ at $z=3$, 
where
$J_{21}=
J(912{\rm \AA})/ 10^{-21}  \rm erg \ s^{-1}cm^{-2}Hz^{-1}str^{-1}$.
For the spherical model the constant flux model 
with $J_{21}=1$ is also considered.
This will change the formation epoch of stars.
Ionization equilibrium is assumed, and 
radiative cooling and UV heating are taken into account.

\subsection{Spherical cloud model} 

First we examine propagation of a galactic wind from a minihalo
with a simple spherical model.
A system of dark matter and gas ($\Omega_b=0.1$) evolves
from a Gaussian density fluctuation
\cite{BSS}.
Star formation criterion is set as
$    \rho_b  > 1.67 \times 10^{-24} \rm g \ cm^{-3}$, and
$T  < 10^4 \rm K.$
Stars are formed with the timescale, $\tau_{SF}$.

An expanding shell due to supernova explosions propagates 
into the expanding region, being accumulated all gas which is collapsing
onto the minihalo.
The expansion is accelerated by the pressure gradient.
The expanding velocity of the shell when it is propagating into 
the IGM region is 
 $V_{exp} \simeq 150 - 300 $km/s  ($\tau_{SF}=10^7 - 3 \times 10^7$yr). 
In this spherical model the metallic gas is  
confined in the hot cavity region. 
When $\tau_{SF}$ is long such as $10^8$yr, supernovae 
do not produce
an expanding shell. 
Lasting period of the star formation is short  $\sim (3-4) \tau_{SF}$. 
The stellar component ($M_{\ast} \lsim (0.03-0.05)M_g$) 
 distributes at  $R_{\ast} \lsim 3-10$ kpc
(depending on the initial density fluctuation). 
 Average metallicity of the system is
  $ Z= M_Z/M_g \simeq   
 0.06-0.005 Z_{\odot} $
at $z \sim 3$.

\begin{figure}
\centerline{\hbox{
\psfig{figure=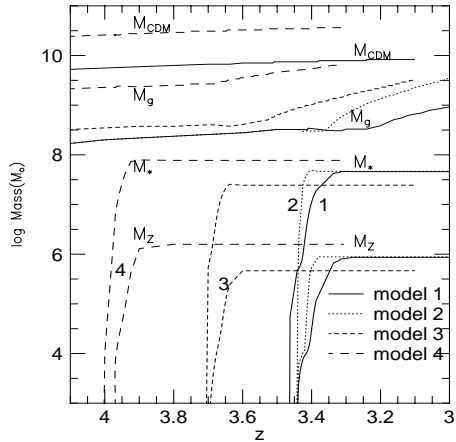,width=2.5in,height=2.4in,bbllx=72bp,bblly=140bp,bburx=555bp,bbury=652bp,clip=}
}}
\caption{Redshift evolution of metallic gas mass($M_Z$), stellar mass
($M_{\ast}$), gas mass($M_g$), and CDM mass ($M_{CDM}$)
for model 1 ($J_{21}=1$, $\tau_{SF}=3 \times 10^7$yr, $r_c=90$kpc),
 model 2 ($J_{21}=1$, $\tau_{SF}= 10^7$yr, $r_c=90$kpc),
 model 3 ($J_{21}\sim 0.7$, $\tau_{SF}= 10^7$yr, $r_c=90$kpc), and
 model 4 ($J_{21}\sim 0.7$, $\tau_{SF}= 10^7$yr, $r_c=150$kpc),
where $r_c$ is the core radius of the CDM distribution.  
The gas mass increases because the expanding shell sweeps the IGM.}
\end{figure}

\subsection{Grid geometry model } 

To examine how metals expand in a space
where voids (low density region) and gas sheets or wall (high density region)
exist,
we consider a toy model with grid geometry of  walls and  voids.
Here we consider a
$2h^{-1}$ Mpc box (comoving) with $64^3$ mesh and $64^3$ CDM particles.
Walls separate 8 void regions in the simulation box.
The void under-density is assumed as $\delta \rho_V= 0.80 $
and 
the wall over-density, $\delta \rho_W= 1.62 $. 
 The wall width is put as $ 0.25h^{-1}$ Mpc.
A galaxy ($M_{\ast}=5 \times 10^{10} M_{\odot}$)
which is put near the center 
has supernova explosions at $z=4$ to spread out metallic gas into the IGM.

Shocked hot gas shell following cool and dense  gas shell 
expands in the IGM. 
The wall 
prevents the shell expansion 
which propagates toward the voids 
with the expanding velocity, $\sim 200$km/s, at $z=2.75$.
Most metallic gas is accumulated in the dense  shell, where 
CIV number density is high 
as well as HI number density.
The unperturbed walls are not metal enriched.
The IGM metallicity is almost uniform from the central region to the dense shell.
Beyond the shell the metallicity decreases with decreasing gas density.

\subsection{3D IGM model in  CDM model } 

\begin{figure}
\centerline{\hbox{
\psfig{figure=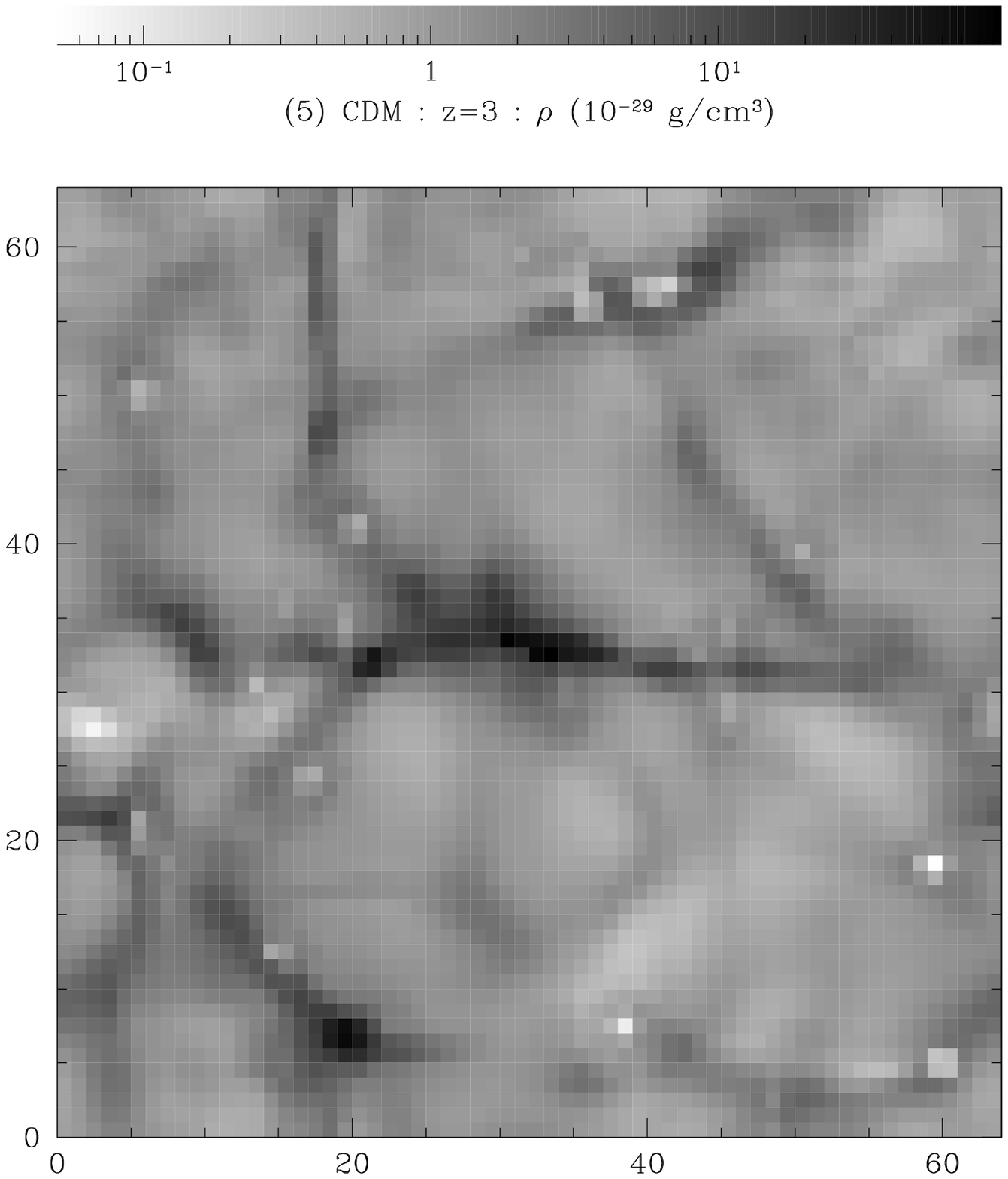,width=2.5in}
\psfig{figure=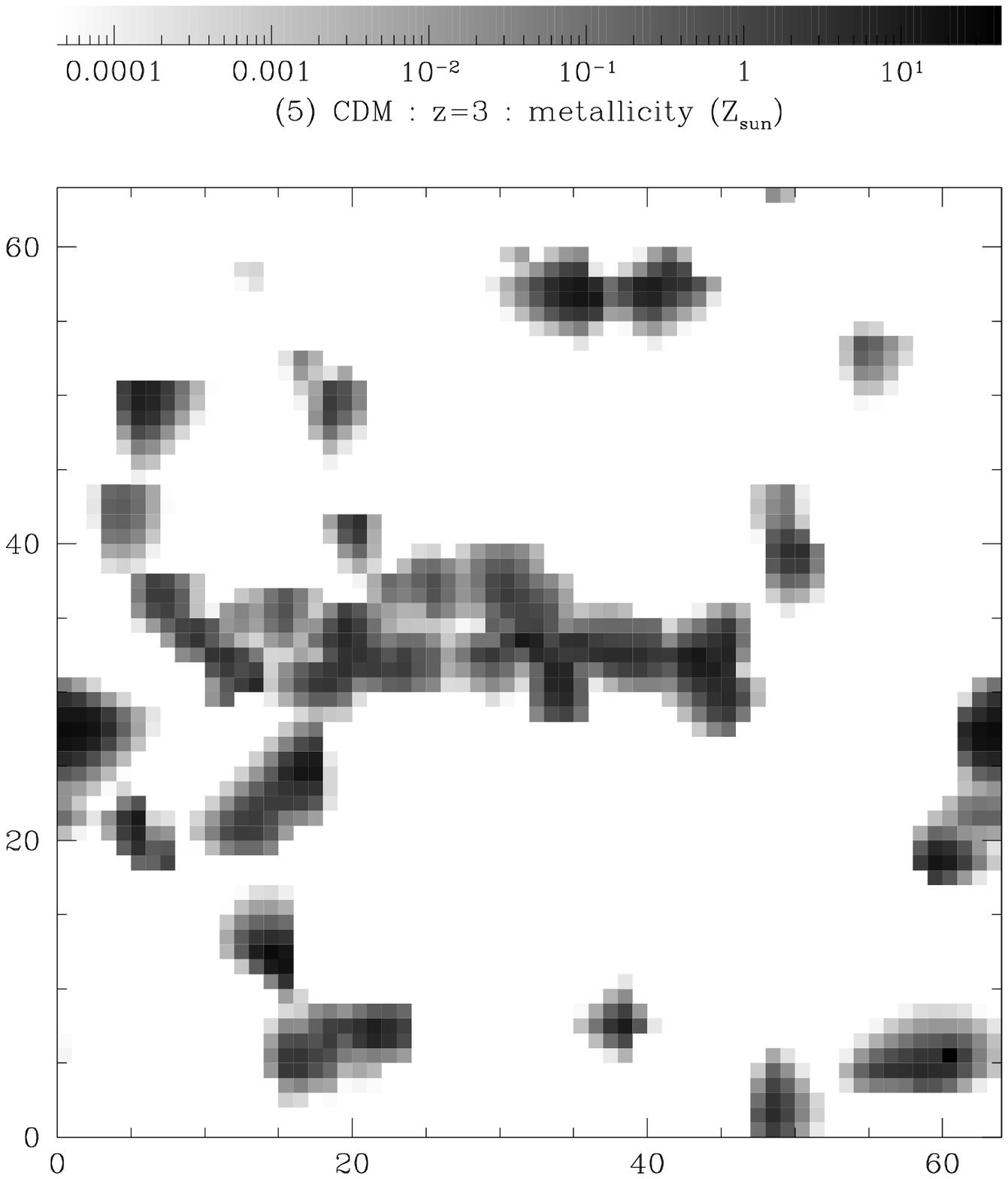,width=2.5in}
}}
\caption{Snapshot of the gas density distribution (left) 
and the metallicity distribution
(right) at $z=3$ in 3D IGM model.}
\end{figure}

We have performed 3D simulations in the CDM model
to examine metal distribution of the IGM.  
We take a $20 h^{-1}$Mpc (comoving) box with
$64^3$ mesh and $64^3$
CDM particles. 
We assume
$\Omega_b=0.052 $,  
and
$\sigma_8= 1.01 $ for the CDM model. 
Galaxies are assumed to form at a cell where 
$\rho_{tot} > 5 \rho_{crit}$, $\rho_b > \bar{\rho}_b$, 
$\nabla \mbox{\boldmath $v$} <0$,
and $T < 2 \times 10^5$K,
and all gas in the cell is converted to stars. 
(e.g. \cite{cen}.)
Supernovae are assumed to explode soon after galaxies are formed.

First we find that supernova feedback 
 acts to suppress galaxy formation near galaxies.
The galaxy formation rate (mass per unit redshift) 
becomes nearly constant at $z <10$, contrary to a model
without supernova feedback, which rate show a peak  around
$z \sim 6$.

The IGM is heated by supernovae 
during $z \approx 5-12$, and 
by gravitational collapse at $z \lsim 5$, in average. 
The metallicity distribution mostly traces the high density regions
where galaxies are formed and metals are produced.
Some low density regions caused by galactic winds and bulk motion
are found to have high metallicity.
The metallicity is less than 1/100 $Z_{\odot}$ at more than 90 \% region
at $z=3$.
Supernova explosions produce two trends in plots of the metallicity
vs. the gas density.
The metallicity keeps high near galaxies with wide range of the gas density.
On the other hand,
expanding gas towards void regions makes a trend of lower metallicity with
lower density.
These trends are clearly seen in the toy grid model in \S 2.2 and
were not found by models without supernova feedback
\cite{GO}.
The CIV column density, N(CIV), spreads widely 
for a fixed HI column density,
because of wide range of the metallicity.
At $z=3$ 
all $  10^{15} \leq $ N(HI)$ <10^{17} \rm cm^{-2}$  clouds seem to have
   N(CIV)$ \geq 10^{12} \rm cm^{-2},$  
similarly to observations \cite{SC96}.

\section{DISCUSSION }

The results of these models show that: 
(1) galactic winds from minihalos and  galaxies 
carry metallic gas to the IGM but are not powerful enough to
pollute all the IGM. 
Even at $z=0$ only 30 \% region 
has metallicity larger than $10^{-2}Z_{\odot}$;
and 
(2) CIV absorption lines
would be associated at high density regions.
The CIV column densities and metallicity 
are likely to  show similar tendency to the observations.

These models are still preliminary and we need more
improvements.
For example,
simulations with higher spatial resolution would make a change of
galaxy formation rate, because 
the formation criterion might allow to form more
galaxies but increased number of supernovae would prevent
succeeding formation of galaxies.
The change of the galaxy formation rate will affect
the precise metallicity distribution.
Improved models are being planed to examine, which will give us
more realistic information on the metallicity of the IGM.


%
\vfill
\end{document}